# Variational Approach to Necessary and Sufficient Stability Conditions for Inviscid Shear Flow

Makoto Hirota[1], Philip J. Morrison[2], Yuji Hattori[1]
[1]Institute of Fluid Science, Tohoku University, Sendai 980-8577, Japan,
[2]Institute for Fusion Studies, University of Texas at Austin, Austin, TX78712, USA
E-mail of corresponding author: hirota@dragon.ifs.tohoku.ac.jp

**ABSTRACT**
A necessary and sufficient condition for linear stability of inviscid parallel shear flow is formulated by a novel variational method, where the velocity profile is assumed to be monotonic and analytic. Unstable eigenvalues of the Rayleigh equation are shown to be associated with positive eigenvalues of a certain selfadjoint operator. The stability is therefore simply determined by maximizing a quadratic form, which is theoretically and numerically more tractable than directly solving the Rayleigh equation. This variational approach is based on the Hamiltonian nature of the inviscid fluid and will be applicable to other hydrodynamic stability problems.

## 1. Introduction

Linear stability conditions for shear flow are generally difficult to derive theoretically, even for the inviscid fluid model. Since the eigenvalue problem for the linearized fluid equation is infinite-dimensional and non-selfadjoint, it is mathematically hard to prove instability for a given equilibrium flow profile (unless the rare analytical solution exists). Although several variational approaches [1-4] have been proposed in the context of Hamiltonian mechanics, it is well-known that they give only sufficient conditions for stability.

The present work revisits the stability of inviscid parallel shear flow and considers the Rayleigh equation [5]. For this classical problem, we show that the variational approach can be improved so as to give a *necessary and sufficient* condition. Restricting our consideration to monotonic and analytic velocity profiles, we present a quadratic form $Q$ whose positive signature indicates the existence of an unstable eigenmode. Instability can be proven by finding some test function that makes $Q$ positive, which is analytically and numerically feasible without knowing the rigorous solution of the equation.

## 2. Variational stability criterion

Specifically, consider the linear stability of inviscid parallel shear flow $\vec{U} = (0, U(x))$ on a 2D domain $[-L, L] \times [-\infty, \infty]$ bounded by two walls at $x = \pm L$. By introducing the stream function of the disturbance as $\phi(x) e^{ik(y-ct)} + \text{c.c.}$ with a complex phase speed $c$ and a wavenumber $k > 0$, the stability is determined by Rayleigh's equation [5];

$$(c-U)(\phi''-k^2\phi)+U''\phi=0, \qquad (1)$$
$$\phi(-L)=\phi(L)=0,$$

where the prime (') indicates the $x$ derivative. If this equation has a nontrivial solution for $c$ with a positive imaginary part, $\operatorname{Im} c > 0$, the shear flow is spectrally unstable due to an exponentially growing eigenmode.

Rayleigh's necessary condition for instability [5] is that the velocity profile $U(x)$ must have at least one inflection point; i.e., an $x = x_I$ at which $U''(x_I) = 0$.

Arnold [1] improved this stability condition by using a variational approach. He showed that the shear flow $U(x)$ is stable if the quadratic form,

$$Q = \int_{-L}^{L} \xi[(U-U_I)+U''G]U''\xi dx. \qquad (2)$$

is either positive or negative definite (see [4]), where $U_I = U(x_I)$ and the linear operator $G$ denotes a convolution integral defined by $G^{-1}\phi = -(\phi''-k^2\phi)$. Physically, (2) represents the second variation of the energy with respect to the fluid displacement $\xi$.

In this work, we further improve this variational criterion by introducing the following assumptions.

**Assumption (A1)**: *the velocity profile $U(x)$ is an analytic, bounded, and strictly monotonic function. Also, if $U''(x_I) = 0$ at $x = x_I$, then $U'''(x_I) \neq 0$.*

If $U(x)$ satisfies (A1) and has only one inflection point $x_I$, then either $U''(U-U_I) \geq 0$ or $U''(U-U_I) \leq 0$ holds for all $x$. In the former case, the quadratic form (2) is positive definite and the flow is stable, which is also known as Fjortoft's theorem [6].

Thus, our concern is with the latter case, for which we present the following theorem.

**Theorem 1.** *Let $U(x)$ satisfy (A1) and have one inflection point $x = x_I$. When $U''(U-U_I) \leq 0$ for all $x$, the shear flow is spectrally stable if and only if the quadratic form (2) is not positive for all $\xi \in L^2$.*

We can generalize this theorem to the case of multiple infection points as follows.

**Theorem 2.** *Let $U(x)$ satisfy (A1) and have inflection points $x_{I_n}$, $n=1,2,...,N$. The shear flow is spectrally stable if and only if the quadratic form,*

$$Q = \nu \int_{-L}^{L} \xi \prod_{n=1}^{N} [(U-U_{I_n})+U''G]U''\xi dx. \qquad (3)$$

*is not positive for all $\xi \in L^2$, where $U_{I_n} = U(x_{I_n})$ and either $\nu = 1$ or $\nu = -1$ is chosen such that*

$$\nu U'' \prod_{n=1}^{N}(U-U_{I_n}) \leq 0 \quad \text{for all } x.$$

We remark that the functional space $L^2$ for $\xi$ may be replaced by other Hilbert spaces satisfying certain conditions. Actually, it is technically convenient to maximize $Q$ with respect to $w = U''\xi \in L^2$ rather than $\xi \in L^2$. Then, the number of positive eigenvalues of the selfadjoint operator $H$, defined by $Q = \int_{-L}^{L} w H w \, d$, corresponds to the number of unstable eigenvalues of the Rayleigh equation.

### 3. Numerical tests

Here, we exhibit a few numerical results to illustrate how our method works in practice. For each velocity profile $U(x)$, we compare the results of two different numerical codes: one code solves Rayleigh's equation (1) directly to search for complex eigenvalues $c$, while the other solves for the eigenvalues $\lambda_1, \lambda_2, ...$ of the selfadjoint operator $H$ in descending order.

The first example is

$$U(x) = x + 5x^3 + 1.62\tanh[4(x-0.5)] \quad (4)$$

for $x \in [-1,1]$, which is studied by Balmforth and Morrison [7]. This flow has three inflection points and is unstable only for finite wavenumber $k_3 < k < k_2$, unlike the usual Kelvin-Helmholtz instability. As shown in Fig. 1, the maximum eigenvalue $\lambda_1$ (the dashed line) smoothly changes its sign at $k = k_2, k_3$ and becomes positive for $k_3 < k < k_2$.

The second example is

$$U(x) = x - 0.02 + \sin[8(x-0.02)]/16 \quad (5)$$

for $x \in [-1,1]$, which has five inflection points. As shown in Fig. 2, three unstable eigenvalues emerge at $k_1, k_3, k_5$ with different phase speeds $U_{I1}, U_{I3}, U_{I5}$, respectively. The three eigenvalues $\lambda_1, \lambda_2, \lambda_3$ also become positive at these critical wavenumbers.

These numerical results confirm that the signature of $Q$ indeed predicts the existence of unstable eigenvalues ($\text{Im}\, c > 0$).

### 4. Concluding remarks

We remark that our stability criterion can reproduce the earlier results of the Nyquist method [7] and the perturbation analysis of the neutral modes [8]. Our variational approach is advantageous in that we can prove instability by merely finding a test function (in a certain function space) that makes the quadratic form $Q$ positive. For the purpose of determining stability, the variational problem (i.e., the signature of $Q$) can be solved more quickly and accurately than the Rayleigh equation (a non-selfadjoint eigenvalue problem with singularity). This variational approach is expected to be applicable to other hydrodynamic stability problems and to be of practical utility in other complicated problems.

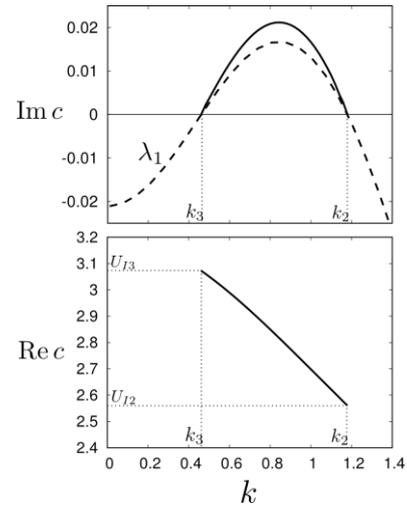

Fig. 1 Growth rate ($\text{Im}\, c$) and phase speed ($\text{Re}\, c$) versus wavenumber $k$ for the shear flow (4).

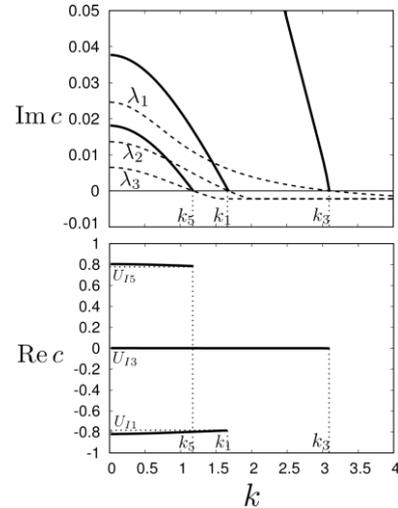

Fig. 2 Growth rate ($\text{Im}\, c$) and phase speed ($\text{Re}\, c$) versus wavenumber $k$ for the shear flow (5).

**References**
[1] V. I. Arnold, *Mathematical Methods of Classical Mechanics*, Springer, Berlin (1978).
[2] D. D. Holm *et al.*, Phys. Rep. **123** (1985), 1.
[3] P. J. Morrison, Rev. Mod. Phys. **70** (1998), 467.
[4] N. J. Balmforth and P. J. Morrison, *Hamiltonian Description of Shear Flow*, in Large-Scale Atmosphere-Ocean Dynamics II, eds. J. Norbury and I. Roulstone (Cambridge, Cambridge, 2002), 117.
[5] J. W. S. Rayleigh, Proc. Lond. Math. Soc. **9** (1880), 57.
[6] R. Fjortoft, Geofys. Publ. **17** (1950), 1.
[7] N. J. Balmforth and P. J. Morrison, Stud. Appl. Math. **102** (1999), 309.
[8] W. Tollmien, Nachr. Wiss Fachgruppe, Gottingen, Math. Phys., **1** (1935), 79.